\documentclass[conference]{IEEEtran}
\IEEEoverridecommandlockouts

\usepackage{cite}
\usepackage{amsmath,amssymb,amsfonts}
\usepackage{algorithmic}
\usepackage{graphicx}
\usepackage{textcomp}
\usepackage{xcolor}
\usepackage{balance}
\usepackage{booktabs}
\usepackage{url}
\usepackage{siunitx}
\usepackage{multirow}
\usepackage{amsmath}
\usepackage{float}
\usepackage{pifont}
\usepackage{makecell}
\usepackage[utf8]{inputenc}
\usepackage[T1]{fontenc}
\usepackage{hyperref}

\def\BibTeX{{\rm B\kern-.05em{\sc i\kern-.025em b}\kern-.08em
    T\kern-.1667em\lower.7ex\hbox{E}\kern-.125emX}}

\begin{document}
\title{Attacking Voice Anonymization Systems with Augmented Feature and Speaker Identity Difference}

\author{
\IEEEauthorblockN{
Yanzhe Zhang$^1$,
Zhonghao Bi$^2$,
Feiyang Xiao$^1$,
Xuefeng Yang$^1$,
Qiaoxi Zhu$^3$,
Jian Guan$^1$$^\ast$
}
\IEEEauthorblockA{$^1$Group of Intelligent Signal Processing, College of Computer Science and Technology, Harbin Engineering University, China}
\IEEEauthorblockA{$^2$Faculty of Computing, Harbin Institute of Technology, China}
\IEEEauthorblockA{$^3$Acoustics Lab, University of Technology Sydney, Australia}
}

\maketitle

\renewcommand{\thefootnote}{\fnsymbol{footnote}}
\footnotetext{$^\ast$Corresponding author}
\renewcommand{\thefootnote}{\arabic{footnote}}

\begin{abstract}
This study focuses on the First VoicePrivacy Attacker Challenge within the ICASSP 2025 Signal Processing Grand Challenge, which aims to develop speaker verification systems capable of determining whether two anonymized speech signals are from the same speaker. However, differences between feature distributions of original and anonymized speech complicate this task. To address this challenge, we propose an attacker system that combines Data Augmentation enhanced feature representation and Speaker Identity Difference enhanced classifier to improve verification performance, termed DA-SID. Specifically, data augmentation strategies (i.e., data fusion and SpecAugment) are utilized to mitigate feature distribution gaps, while probabilistic linear discriminant analysis (PLDA) is employed to further enhance speaker identity difference. Our system significantly outperforms the baseline, demonstrating exceptional effectiveness and robustness against various voice anonymization systems, ultimately securing a top-5 ranking in the challenge.
\end{abstract}


\section{Introduction}
The First VoicePrivacy Attacker Challenge~\cite{NT2024attacker}, 
part of the ICASSP 2025 Signal Processing Grand Challenge, aims to develop speaker verification systems that determine whether two anonymized speech signals come from the same speaker, effectively attacking voice anonymization systems. The official baseline~\cite{NT2024attacker} 
adopts ECAPA-TDNN~\cite{desplanques2020ecapa} as feature extractor to obtain speaker feature embeddings and a cosine similarity classifier to measure the similarity between anonymized speech for speaker verification.

However, the significant differences in feature distributions between original and anonymized speech make it challenging for an attacker system to determine whether two anonymized speech signals originate from the same speaker. Additionally, voice anonymization systems reduce the distinction between speakers in anonymized speech, further complicating the attack process for speaker verification.

In this paper, based on the official baseline~\cite{NT2024attacker}, we propose a data augmentation (DA) and speaker identity difference (SID) based attacker system, termed DA-SID. The system utilizes data augmentation techniques, i.e., data fusion \cite{meng2020datafusion} and SpecAugment \cite{park2019specaugment}, thereby enhancing the robustness of feature representation and reducing the feature distribution gap between original and anonymized speech. Additionally, probabilistic linear discriminant analysis (PLDA) \cite{kenny2013plda} is employed as the classifier, which effectively separates speaker embeddings by modeling the differences between speakers while accounting for variations within each speaker’s speech. Our DA-SID attacker system achieves significant performance improvements over the official baseline, which was ranked among the top-5 in the First VoicePrivacy Attacker Challenge, highlighting its robustness and competitiveness against state-of-the-art anonymization systems.

\begin{table*}[ht!]
\caption{Performance comparison of our DA-SID attacker system and baseline in terms of EER (\%) against six anonymization systems. $\mathcal{L}_{\text{con}}$ represents contrastive learning. }

\vspace{-4mm}
\begin{center}
\resizebox{0.86\textwidth}{!}{
\begin{tabular}{c c c c c c c c c c c}
  \toprule
  \multirow{2.5}{*}{\makecell{Anonymization System}} & \multirow{2.5}{*}{\makecell{Attacker System}} & \multirow{2.5}{*}{SpecAugment} & \multirow{2.5}{*}{$\mathcal{L}_{\text{con}}$} & \multicolumn{3}{c}{EER on dev-clean subset} & \multicolumn{3}{c}{EER on test-clean subset} & \multirow{2.5}{*}{Total Average EER} \\
  \cmidrule(lr){5-7} \cmidrule(lr){8-10}
  & & & & Female & Male & Average & Female & Male & Average & \\
  \midrule
  \multirow{2}{*}{B3} & Baseline & \ding{55} & \ding{55} & 28.43 & 22.04 & 25.24 & 27.92 & 26.72 & 27.32 & 26.28 \\
  & \textbf{DA-SID} & \textbf{\ding{51}} & \textbf{\ding{51}}
  & \textbf{25.98} & \textbf{21.12} & \textbf{23.55} & \textbf{27.33} & \textbf{21.60} & \textbf{24.47} & \textbf{24.01} \\
  \midrule
  \multirow{2}{*}{B4} & Baseline & \ding{55} & \ding{55} & 34.37 & 31.06 & 32.71 & 29.37 & 31.16 & 30.26 & 31.49 \\
  & \textbf{DA-SID} & \textbf{\ding{51}} & \textbf{\ding{51}} & \textbf{27.68} & \textbf{23.30} & \textbf{25.49} & \textbf{20.26} & \textbf{22.27} & \textbf{21.26} & \textbf{23.38} \\
  \midrule
  \multirow{2}{*}{B5} & Baseline & \ding{55} & \ding{55} & 35.82 & 32.92 & 34.37 & 33.95 & 34.73 & 34.34 & 34.36 \\
  & \textbf{DA-SID} & \textbf{\ding{55}} & \textbf{\ding{55}} & \textbf{32.53} & \textbf{27.80} & \textbf{30.16} & \textbf{28.51} & \textbf{26.45} & \textbf{27.48} & \textbf{28.82} \\
  \midrule
  \multirow{2}{*}{T8-5} & Baseline & \ding{55} & \ding{55} & 39.63 & 40.84 & 40.24 & 42.50 & 40.05 & 41.28 & 40.76 \\
  & \textbf{DA-SID} & \textbf{\ding{51}} & \textbf{\ding{55}} & \textbf{26.42} & \textbf{28.07} & \textbf{27.24} & \textbf{26.07} & \textbf{23.64} & \textbf{24.85} & \textbf{26.05} \\
  \midrule
  \multirow{2}{*}{T12-5} & Baseline & \ding{55} & \ding{55} & 43.32 & 44.10 & 43.71 & 43.61 & 41.88 & 42.75 & 43.23 \\
  & \textbf{DA-SID} & \textbf{\ding{55}} & \textbf{\ding{55}} & \textbf{32.39} & \textbf{29.18} & \textbf{30.78} & \textbf{27.55} & \textbf{26.72} & \textbf{27.14} & \textbf{28.96} \\
  \midrule
  \multirow{2}{*}{T25-1} & Baseline & \ding{55} & \ding{55} & 42.65 & 40.06 & 41.36 & 42.34 & 41.92 & 42.13 & 41.75 \\
  & \textbf{DA-SID} & \textbf{\ding{55}} & \textbf{\ding{55}} & \textbf{35.25} & \textbf{31.37} & \textbf{33.31} & \textbf{33.39} & \textbf{32.27} & \textbf{32.82} & \textbf{33.07} \\
  \bottomrule
\end{tabular}    
}
\end{center}
\label{tab:result}
\vspace{-\baselineskip}
\vspace{-4mm}
\end{table*}

\section{Proposed Attacker System}
\label{sec:description}
\subsection{Data Augmentation Enhanced Feature Representation}
\label{sec:da}
Anonymization systems introduce significant distribution gaps between speaker features before and after anonymization. To mitigate these gaps, we employ data fusion \cite{meng2020datafusion} strategy that combines original dataset $\mathcal{D}_{\text{orig}}$ with the corresponding anonymized dataset $\mathcal{D}_{\text{anon}}$ to obtain the fused dataset $\mathcal{D}_{\text{fused}}$:
\begin{equation}
\label{eq:1}
    \mathcal{D}_{\text{fused}} = \mathcal{D}_{\text{orig}} \cup \mathcal{D}_{\text{anon}}.
\end{equation}
By jointly learning from the fused dataset, the relation between speaker features before and after anonymization is captured, effectively mitigating distribution differences.

Meanwhile, to further enhance the robustness of the feature representation, the SpecAugment \cite{park2019specaugment} strategy is also employed in the extraction of the speaker feature embedding $\mathbf{e}$, as
\begin{equation}
\label{eq:2}
    \mathbf{e} = \mathcal{F}(\mathbf{X} \odot \mathbf{M}_{tf}),
\end{equation}
where $\mathbf{X}$ represents the log-Mel feature of the input speech, $\mathbf{M}_{tf}$ denotes the time-frequency masking matrix applied by SpecAugment, and $\odot$ indicates element-wise multiplication. $\mathcal{F}(\cdot)$ represents the ECAPA-TDNN feature extractor, optimized using the additive angular margin loss \cite{xiang2019margin}. For anonymization systems that retain more speaker information (i.e., B3 and B4), a contrastive loss is equally incorporated to enhance the feature representation, as in \cite{wang2024speaker}. 

\subsection{Speaker Identity Difference Enhanced Classifier}
To enhance speaker identity difference, we employ PLDA \cite{kenny2013plda} trained on an anonymized dataset as the classifier. Specifically, the similarity score between two speaker feature embeddings $\mathbf{e}_i$ and $\mathbf{e}_j$ ($i \ne j$) is calculated in PLDA based on the log-likelihood:
\begin{equation}
    s(\mathbf{e}_i, \mathbf{e}_j) = \log p(\mathbf{e}_i, \mathbf{e}_j \mid H_0) - \log p(\mathbf{e}_i, \mathbf{e}_j \mid H_1),
\label{plda:sim}
\end{equation}
where $p(\mathbf{e}_i, \mathbf{e}_j\!\mid\!H_0)$ and $p(\mathbf{e}_i, \mathbf{e}_j\!\mid\!H_1)$ denote the probabilities under the hypotheses $H_0$ and $H_1$, which represent that $\mathbf{e}_i$ and $\mathbf{e}_j$ are from the same speaker or different speakers, respectively.

The optimization of the PLDA classifier increases the similarity score for anonymized speaker feature embeddings belongings to the same speaker while decreasing it for embeddings from different speakers, thereby emphasizing distinctions in speaker identity. By leveraging PLDA as the classifier, our system enhances speaker identity difference during verification, thereby improving the overall attack performance.

\section{Results}
\begin{table}[t]
\caption{Ablation study on the impact of data augmentation (DA) and speaker identity difference (SID) on EER (\%).}
\vspace{-5mm}
\begin{center}
\resizebox{0.85\linewidth}{!}{
\begin{tabular}{l c c c c c c}
  \toprule
  Method & B3 & B4 & B5 & T8-5 & T12-5 & T25-1 \\ 
  \midrule
  \textbf{DA-SID} & \textbf{24.04} & \textbf{23.42} & \textbf{28.82} & \textbf{26.05} & \textbf{28.96} & \textbf{33.07} \\
  \multicolumn{1}{@{\hspace{10pt}}l}{w/o DA} & 26.06 & 24.41 & 29.22 & 26.96 & 29.70 & 33.75 \\
  \multicolumn{1}{@{\hspace{10pt}}l}{w/o SID} & 25.54 & 27.10 & 33.10 & 41.07 & 37.20 & 38.93 \\
  \multicolumn{1}{@{\hspace{10pt}}l}{w/o DA \& SID} & 26.28 & 31.49 & 34.36 & 40.76 & 43.23 & 41.75 \\
  \bottomrule
\end{tabular}
}
\end{center}
\label{tab:ablation}
\vspace{-\baselineskip}
\vspace{-5mm}
\end{table}

\noindent \textbf{Dataset and Metric:} Following~\cite{NT2024attacker}, our experiments are conducted on the LibriSpeech dataset \cite{panayotov2015librispeech} and the equal error rate (EER) is used as the performance metric, where a lower EER indicates better attack performance.

\noindent\textbf{Performance Comparison:} The attack performance on six anonymization systems is presented in Table~\ref{tab:result}. The proposed DA-SID system consistently outperforms the official baseline across all anonymization systems. Specifically, for T8-5, DA-SID achieves a remarkable EER reduction of 14.71\%, establishing it as the most effective attacker system for T8-5 among all the attacker systems in the challenge~\cite{NT2024attacker}.

\noindent \textbf{Ablation Study: }To further evaluate the effects of DA representation  and SID-enhanced classifier, we conduct an ablation study, as shown in Table~\ref{tab:ablation}. Here DA is optimized with additive angular margin loss. The results show a significant degradation in EER when either of these components is removed (denoted as w/o), confirming that both DA representation and SID-enhanced classifier are effective in improving attack performance. Furthermore, the results indicate that SID-enhanced classifier provides a greater performance boost compared to DA representation, highlighting it as a more impactful design.

Note that, as the distributions of the anonymized speech pairs for T10-2 anonymization system are mismatched, DA-SID cannot be applied to T10-2 in the challenge. To address this limitation, we employ a pretrained model TitaNet-Large \cite{koluguri2022titanet} to extract speaker embeddings, combined with a cosine similarity classifier for speaker verification. Our method achieves an EER of 32.23\%, notably outperforming the official T10-2 baseline system of 41.10\%.

\section{Conclusion}
\label{sec:conc}
We proposed a DA-SID attacker system to address the significant feature distribution differences between original and anonymized speech. By integrating data augmentation with probabilistic linear discriminant analysis, our system effectively mitigates distribution gaps and enhances speaker identity difference. Experimental results demonstrate that our system outperforms the official baseline, which was ranked among the top-5 systems in the challenge.

\balance
\bibliographystyle{IEEEtran}
\bibliography{mybib}

\end{document}